\documentclass[11pt]{article}

\usepackage[margin=1in]{geometry}
\usepackage{amsmath,amssymb}
\newcommand{\Id}{\mathbf{I}}
\usepackage{graphicx}
\usepackage{booktabs}
\usepackage{hyperref}
\usepackage[numbers,sort&compress]{natbib}
\usepackage{xcolor}

\providecommand{\doi}{}
\renewcommand{\doi}[1]{\href{https://doi.org/#1}{doi:#1}}
\usepackage{array}

\title{Scouting Super Resolution: from HLT-Level Reconstruction to
  Offline-like Quality} \author{Maurizio Pierini\\[2pt]\small European
  Organization for Nuclear Research (CERN), Geneva, CH-1211,
  Switzerland} \date{\today}

\begin{document}
\maketitle

\begin{abstract}
Data scouting, introduced by the CMS collaboration in 2011, is a
technique to extend the reach of the experiment beyond what full-event
storage bandwidth allows: rather than the complete event record, only
the physics objects already reconstructed by the High-Level Trigger
(HLT) are persisted, which makes it affordable to record
$\mathcal{O}(10)$ times more data than the standard physics
stream. The price for this rate increase is in experimental
resolution: online reconstruction has typically a lower resolution and
a worse response, mostly due to shortcuts taken to speed up the
reconstruction and to the use of suboptimal calibration constants. For
this reason, scouting data has mostly been used when the resolution
loss is negligible in absolute terms (muons) or with respect to the
offline resolution (jets). In any case, the need to derive specific
online-to-offline corrections has been a limiting factor. In this
paper, we show that a regression trained on a given object and the
surrounding particle cloud can bring the gap between online and
offline resolution within the typical uncertainty band of the offline
data. This procedure could be applied to the scouting monitoring
stream, the subset of scouting data also reconstructed offline. We
demonstrate this on CMS Run-1 Open Data, reprocessed through paired
HLT and offline passes over the same events. One architecture, applied
unchanged to electrons and jets, suppresses residual biases by one to
two orders of magnitude on various quantities (kinematics, isolation,
energy composition) and achieves a resolution improvement. Measured
differentially in $p_T$, the corrected objects fall inside the offline
calibration uncertainties CMS quotes for this dataset, where the
uncorrected HLT objects lie outside them: a corrected object can be
treated with the standard, centrally provided calibration, bringing
scouting data much closer to offline data for practical purposes.
\end{abstract}

\section{Introduction}
\label{sec:intro}

Data scouting is a strategy developed by the Compact Muon Solenoid (CMS)
experiment~\citep{cms:detector} during
Run~1 and expanded
through Run~2 and Run~3 to extend the physics reach of the experiment
beyond what full-event storage bandwidth would otherwise allow
\citep{cms:scouting2024}, first deployed in a search for narrow dijet
resonances~\citep{cms:scouting2016}. Instead of writing out the complete raw and
reconstructed event record, a scouting stream persists only the physics
objects already reconstructed by the High-Level Trigger
(HLT)~\citep{cms:trigger2017} itself (tracks,
electrons, muons, jets, and the underlying trigger objects), discarding both
the raw detector data and the fuller, more precise
reconstruction that an offline pass would otherwise produce. This trades
per-object information content for event rate: because a scouting record
is far smaller than a full event, CMS can afford to write scouting streams
at rates $\mathcal{O}(10)$ times higher than the full physics
stream, which in turn allows searches for signatures (low-mass
resonances, soft unconventional signals, trigger-level anomaly
searches~\citep{l1anomaly2022})
that would otherwise be prescaled away or never triggered on at all
\citep{cms:scouting2024}. The same paradigm has been adopted by the other
Large Hadron Collider (LHC) experiments: ATLAS performs trigger-level
analyses~\citep{atlas:tla2018} and LHCb records its
physics output through the Turbo stream~\citep{lhcb:turbo2015}, in which
the online reconstruction is the one used for analysis. The cost of that higher rate is resolution: HLT
reconstruction is built under tight latency constraints, uses coarser or
regional information, and in general skips the iterative,
calibration-heavy refinements that offline reconstruction applies before
an object's kinematics and identification variables are considered final.
An object in a scouting stream is therefore a systematically lower-
fidelity version of the same object an offline analysis would see: not
merely noisier, but missing entire categories of information.

This resolution loss is not uniform across object types and time, and
the non-uniformity is what has shaped where scouting has been used so
far \citep{cms:scouting2024} (mostly jets and muons). For muons the
degradation is small and, more importantly, stable in time. For jets
the resolution loss is bigger than for muons, but its physics impact
is mitigated by the offset of the sizeable offline jet
resolution. 

The situation is more complicated for electrons and photons: the
crystals of the electromagnetic calorimeter (ECAL) lose transparency as they accumulate radiation dose, so the
calorimeter response drifts continuously over a run, and partially
improve during beam-off time. The corrections that undo this drift are
derived from the very data they must be applied to: events are parked
at the CERN Tier-0, the calibration constants are extracted from them,
and the reconstruction is then run with constants matched to the
conditions of that same data. The HLT cannot do this (at trigger time
the constants describing the current state of the detector do not yet
exist), so online reconstruction necessarily runs with older ones.
The resolution loss caused by this calibration difference was large in
Run~1; it has been reduced in Runs~2 and~3 by refreshing the HLT
calibrations more frequently, but it has not been eliminated. The same
logic applies to the absolute scale and to the crystal-by-crystal
inter-calibration, which are set in situ from $Z\to e^+e^-$ decays,
from the $\phi$-symmetry of the energy deposits, from
$\pi^0/\eta\to\gamma\gamma$ decays, and from the $E/p$ ratio of
isolated electrons from $W$ and $Z$
decays~\citep{cms:ecal7tev,cms:egamma2021}: all of these reach their
target precision only once a substantial fraction of a run's
luminosity has been collected. No amount of additional online
computing power closes this part of the gap, since the limiting factor
is calibration statistics accumulated over time, not processing
latency.

We show in this paper that a regression processing the object to be
corrected together with the particle cloud around it (the
particle-flow candidates \citep{cms:pflow2017} the object is built
from or surrounded by, which a scouting stream persists alongside the
object's own high-level quantities) can remove most of the
online/offline difference after the data have been taken, recovering
information that a correction acting on the summary quantities alone
cannot reach. 

We demonstrate these on two object types, electrons and jets, for
which substantial differences exist between online and offline
reconstruction.  For electrons there are two distinct sources. The
first is structural: in Run~1 HLT electrons were built from a
Combinatorial Track Finder (CTF) track rather than the Gaussian-Sum Filter (GSF)
track used offline~\citep{gsf2005}, so every tracking-related quantity differs between
the two views by construction. The second is the calibration lag
described above. Jets are the complementary and harder case. A jet is
a composite object, so all online/offline differences specific of a
particle type enter at several levels at once: which particle-flow
candidates were reconstructed, how they were identified, and
consequently not only the momentum scale but the internal energy
composition and the mass. Among the objects a scouting stream records,
jets are the extreme case of object complexity. 

The results are demonstrated using a graph network inspired by
ParticleNet~\citep{qu2020particlenet}, which is architecturally
capable of processing an unordered set of particles of variable length
(a so-called particle cloud). On the other hand, any architecture able
to process such a set (e.g., a transformer) could be used in its
place. That one architecture, applied unchanged, is able to correct
both objects is not a requirement of the method, but it is a
remarkable outcome: the same architecture absorbs a tracking-algorithm
difference in a single well-instrumented object and a composition
difference in a composite one, which is evidence that what it has
learned follows the underlying mechanism by which resolution
deteriorates, rather than a per-object idiosyncrasy.

In production, such a model would act purely as an offline post-processing
step applied to data already on tape, leaving the trigger, the recorded
event content and the online latency budget untouched
(Section~\ref{sec:deployment}). The training sample it needs (the same
physical object seen twice, once as the HLT reconstructs it and once as the
offline reconstruction does) is one CMS already records as part of
scouting operations, in the \emph{scouting monitoring stream}: a small,
randomly sampled subset of scouting events that is additionally written out
in full and reconstructed offline. That subset is drawn from the same
trigger, the same detector conditions and the same calibration lag as the
data the correction would be applied to, so a real deployment could train on
it directly and retrain periodically as conditions evolve.

The monitoring stream is internal to the CMS collaboration and not
part of any public release, so it is not available here. We therefore
build an equivalent paired sample from CMS 2012 Open Data,
reprocessing the same raw events through both an HLT replay and a
standard offline reconstruction pass (Section~\ref{sec:dataset}).  One
consequence of working with Open Data is that, of the two sources of
electron difference identified above, only the first is
accessible. The online conditions of the 2012 HLT menu are not
published, so the HLT replay must be run with the offline global tag
(that is, with offline calibrations), which artificially removes the
calibration-driven part of the difference. What survives is the
structural difference, and it is large enough for the proof of concept
to stand, with the caveat that a real deployment would face both.

\section{Related work}
\label{sec:related}

Learning a mapping between two representations of the same underlying
collision is an established idea in collider physics, though it has not
previously been applied to the online/offline gap that limits scouting.

Closest in spirit is the super-resolution of calorimeter showers.
\citet{kakati2024srcalo} train a graph-based generative model (continuous
normalizing flows with flow matching, on a transformer backbone) to expand
coarse-granularity, noisy calorimeter cells into a fine-granularity,
denoised representation, and demonstrate, including specifically for
single electrons, that this recovers energy resolution and improves
substructure and downstream particle-flow reconstruction relative to the
low-resolution input. The present work asks the analogous question one
processing stage downstream. \citeauthor{kakati2024srcalo} operate on raw
calorimetric hits, upstream of any object reconstruction, recovering
detector-level granularity before a particle is even formed; data scouting
never retains raw hits at all, so only the already-reconstructed HLT
objects and their particle-flow candidates survive. The question here is
therefore whether super-resolution still succeeds one level higher up the
chain, recovering offline-quality \emph{object-level} features with no
access to the underlying hits.

Three further lines of work share the structure of the problem while
differing in what plays the role of the target.
OmniFold~\citep{OmniFold} showed that a reweighting function between
reconstructed and particle-level distributions can be learned directly
from data in an unbinned, high-dimensional setting, rather than
through a fixed set of binned observables. HGPflow~\citep{HGPflow}
learns a hypergraph-based correction that recovers particle-level
properties from the proxy particles produced by particle-flow
reconstruction, directly analogous in spirit to what is done here, but
applied at a different stage of reconstruction and to
calorimeter-level rather than object-level inputs. CMS
FlashSim~\citep{CMSFlashSim} goes in the opposite direction, learning
a generative map from generator-level truth directly to fully
reconstructed analysis objects, bypassing detector simulation and
reconstruction entirely; the problem addressed here is the mirror
image of a piece of that map, going from a lower-fidelity
reconstructed representation to a higher-fidelity one.

What distinguishes the present problem from all of these is that both the
input and the target are \emph{real, data-derived reconstructions of the
same physical object}, rather than a simulation-to-truth or a
reconstruction-to-generator mapping. This removes any reliance on Monte
Carlo modelling accuracy for the training target. The cost is that it
requires a dataset in which both views exist for the same collisions, which
is what the scouting monitoring stream supplies in production. The
reprocessing pipeline used for this work, described in
Section~\ref{sec:dataset}, provides the closest approximation to that
real-time workflow, within the limited information available for CMS Open
Data.

\section{Dataset and Reprocessing Pipeline}
\label{sec:dataset}

We use CMS Open Data record 64 \citep{opendata:record64}: the
\url{/SingleElectron/Run2012B-v1/RAW} primary dataset (PD), runs
194117--194199 (2012 Run B, $\sqrt{s}=8$\,TeV), 183 files,
$\approx 2.1\times 10^{6}$ events. Only luminosity sections certified for
physics analysis are used, via the standard validated-run JSON
\citep{opendata:record1002}. Reconstruction uses release
\texttt{CMSSW\_5\_3\_32} (\texttt{slc6\_amd64\_gcc472}) and global tag
\texttt{FT53\_V21A\_AN6}, run inside the corresponding CMS Open Data
container image \citep{opendata:docker}.

The PD choice is driven by the need to collect a large enough electron
dataset for training: events in the considered PD are collected by
single-electron triggers, so essentially every event contains a
valuable electron. A large fraction of these events also contains at
least one jet recoiling against the electron, on which no explicit
selection is applied. In particular, no hard $p_T$ requirement is
imposed on the jets so that one can build a dataset populating the
entire range of jet $p_T$ normally considered in the analysis. A
hadronic trigger would have given access only to jets of several
hundred GeV.

For each RAW file we run two independent \texttt{cmsRun} jobs over the
\emph{same} lumi-masked event set:
\begin{itemize}
\item \textbf{Offline reconstruction.} The standard
  reconstruction chain used by CMS
  \citep{opendata:raw2aod}, producing output in the standard analysis object data tier.
\item \textbf{HLT replay.} The exact online HLT menu configuration
  active for this run range,
  \url{/cdaq/physics/Run2012/7e33/v2.1/HLT/V12}
  \citep{opendata:record6139}, limited to the reconstruction steps
  required to create the HLT electron and jet collections, without
  applying any Level-1 trigger seed requirement or any other selection. The online
  detector conditions are not available for CMS Open Data. In lack of
  a better proxy, we used the same conditions used for the offline
  reconstruction.
\end{itemize}

Events are selected online by requiring the
\texttt{HLT\_Ele27\_WP80\_v10} trigger path, whose electron
reconstruction sequence is configured to use the CTF for track
reconstruction. Instead, electrons are reconstructed offline using the
dedicated GSF fit~\citep{gsf2005}, including a brem-recovery
machinery. The use of a different tracking algorithm for online and
offline electrons has a direct consequence on the distribution of many
electron-related features and is an example of the kind of
online/offline differences that one wants to learn to correct for.

In both reconstruction steps, jets are clustered from particle flow
(PF) candidates with the anti-$k_T$
algorithm~\citep{Cacciari:2008gp,fastjet},
configured with jet size parameter $R=0.5$. The resulting online jets
are paired against the offline ones through angular matching. 

Both online and offline jets are selected requiring $p_T > 20$\,GeV
and $|\eta| < 2.5$.  Jets with fewer than three charged particles are
discarded, to remove spurious jets as well as single isolated objects
(electrons, muons, etc.)  reconstructed as individual jets. In
addition, jets within $\Delta R < 0.4$ of the electron are ignored, to
avoid double-counting electrons as jets.

As a default, we persist the list of PF candidates, similarly to what
is done with HLT scouting and we use it to build the particle cloud
given as input to the regression network.

In addition we persist the pixel tracks, to complement that cloud with
the charged particles the online reconstruction found at seed level
but never promoted to a candidate.  This should allow to partially
recover the information lost at the HLT, where the tracking algorithm
only runs the first few steps of the iterative
tracking~\citep{cms:tracking2014}. The pixel
vertices and the online particle-flow track collection are kept
alongside them, so that each track can be referred to a vertex and
each candidate to its own fitted track.

HLT and offline electron candidates are paired by supercluster
position, $\Delta R(\eta_{\rm SC}, \phi_{\rm SC}) < 0.1$, taking the
closest offline match per HLT candidate.

For each matched pair we record various quantities on both the HLT and the offline
side: the electron's transverse momentum $p_T$, pseudorapidity $\eta$, azimuth
$\phi$ and energy $E$; the supercluster position $(\eta_{\rm SC},\phi_{\rm SC})$,
kept separately from the electron object's own direction; the shower-shape
variable $\sigma_{i\eta i\eta}$; the hadronic-to-electromagnetic energy ratio
$H/E$; the ECAL, hadron calorimeter (HCAL) and tracker isolation sums; and the track--cluster matching
variables $\Delta\eta_{\rm in}$ and $\Delta\phi_{\rm in}$.

These play three different roles in the regression, summarized in
Table~\ref{tab:coverage}. Six offline quantities are the targets:
the energy $E$, the ECAL isolation, $\Delta\eta_{\rm in}$,
$\Delta\phi_{\rm in}$, $\eta$ and $\phi$. Two further quantities are
reported but not regressed. The predicted transverse momentum is
derived as $p_T=E/\cosh\eta$ from the predicted energy and the HLT
$\eta$, so that it does not add a target the network would have to
keep consistent with the others. At the HLT, the other electron-specific quantities
($p_T$, $\sigma_{i\eta i\eta}$, $H/E$, and the HCAL and tracker isolation sums)
already have offline-like precision, mostly due to the use of the offline global tag
for the online reconstruction. They are not considered regression targets, but
are provided as input to the network as auxiliary information, to model the correlation 
between them and the correction to the target quantities.

The list of auxiliary quantities is extended with eight HLT-only
track-quality variables: the normalized $\chi^2$, the valid, lost and
valid-pixel hit counts, the quality bitmask, the beamspot-relative
$d_0$ and $d_z$, and the charge.  

In addition, event-specific information is given as input to the
regression: event timestamp, average energy $\rho$, number of primary
vertices, and the number of tracks. The timestamp is required to model
possible effects with time. It is computed using the Unix time with
microsecond precision, read directly from the event record. The other
three quantities provide inputs related to the amount of pileup in the
event.

The supercluster position serves only to match the HLT and offline
candidates but it is deliberately excluded from the regression schema,
since the information of he position of the electron in the detector
is already encoded by the other quantities.

HLT and offline jets are paired by the jet axis directly, $\Delta
R(\eta,\phi) < 0.4$, taking the closest offline jet per HLT jet.  The
cone is correspondingly wider than the electron one, since a jet axis
genuinely moves between the two reconstructions by more than a
supercluster does.

For each matched pair we extract, on both sides, the jet $p_T$,
$\eta$, $\phi$, mass, and three particle-flow energy fractions (EF):
photons, neutral hadrons, and charged particles (including electron
and muon), together with the same auxiliary event-specific quantities used
for the electrons: three event-level pileup proxies and the timestamp.

Only the charged and neutral fractions are regressed; the photon
fraction is recovered as $1-{\rm charged}-{\rm neutral}$, since once
normalized it carries no independent information.

The distinguishing input of the models presented here, relative to a
purely high-level-feature regression, is that each object carries the
particle-flow candidates it is made of (jets) or is surrounded by
(electrons). Both objects use one shared cloud schema, so that the two
pipelines cannot drift apart: the top $N=50$ candidates ordered by
descending $p_T$, each described by
\begin{equation}
(p_T,\ \eta,\ \phi,\ \text{4-way one-hot type}),
\end{equation}
where the type categories are positively charged, negatively charged,
neutral hadron, and photon. Charged candidates are split by the sign of
the charge rather than by PF identity, so that PF muons and electrons fold
into the charged categories consistently with the energy-fraction
regrouping above. Objects with fewer than 50 candidates are zero-padded,
and the validity mask used throughout the network is simply whether the
one-hot type bits sum to one: padded slots are excluded from the graph
construction, from every aggregation, and from the pooling.

The two objects differ only in which candidates enter:
\begin{itemize}
\item \textbf{Jets:} the HLT jet's own constituents. 
\item \textbf{Electrons:} every PF candidate within $\Delta R < 0.5$
  of the HLT electron, including the electron's own footprint. Not
  removing the electron is important to provide the information about
  the main object also in the particle-cloud processing step of the
  network, which engineers the high-level features used in the
  regression.
\end{itemize}
The candidate coordinates are additionally supplemented, at training
time. The coordinates of the particles in the cloud are $(\Delta\eta,
\Delta\phi)$ relative to the main object (with $\Delta\phi$ wrapped to
$[-\pi,\pi]$).

Each charged particle additionally carries the longitudinal and
transverse impact parameters $d_z$, $d_{xy}$ with respect to the
leading HLT pixel vertex, and the fit $\chi^2$ per degree of freedom.
This information is zero-padded for neutral
candidates.

The HLT runs a truncated version of the offline iterative tracking
sequence~\citep{cms:tracking2014}, whose iteration~0 is seeded from pixel
tracks. A pixel track with no corresponding charged
particle-flow candidate is therefore a track the trigger reconstructed at
seed level and failed to promote, the closest available proxy for one that
offline reconstruction would recover. We write the 30 highest-$p_T$ such
tracks per object, taking ``unmatched'' to mean no charged candidate within
$\Delta R<0.02$, each as $(p_T,\eta,\phi,d_z,d_{xy},\chi^2/\mathrm{ndof})$.

These are noisy objects, and typically result in lower-quality
offline tracks. Their momentum measurement carries
little information, due to the poor resolution of pixel-only tracking
at high $p_T$. Nevertheless, the information on their presence is
meaningful. We found it to be particularly relevant to regress the
momentum of hard jets. The typical multiplicity of these objects is
small, but it also carries information: a mean of 2.5 unmatched tracks
per jet and 2.6 per electron cone, rising from 2.5 at 30--40\,GeV to
6.3 above 250\,GeV as jets become denser and more collimated, with the
fraction of jets having none falling from 15\% to 5\%. About one
charged particle in three is seen by pixel tracking but never
promoted, and that ratio is remarkably flat in $p_T$.

The two clouds are merged into a single graph rather than processed
separately: the type one-hot is widened to five categories, the fifth
flagging an unpromoted pixel track, and both share the kinematic, angular
and track-quality columns. The merging is found to provide a better exploitation
of the additional information: built in $(\eta,\phi)$,
the first graph puts a lost track in the same neighbourhood as the
candidates around it, so a dense core with several unpromoted tracks in it
presents differently from a clean one. Two separate graphs would discard
precisely that adjacency.

Table~\ref{tab:coverage} summarizes the resulting dataset structure.

\begin{table}[h]
\centering
\caption{Composition of the datasets given as input to the regression tasks.\label{tab:coverage}}
\small
\begin{tabular}{@{}p{3.4cm}p{9.8cm}@{}}
\toprule
Electrons & Fields \\
\midrule
Particle cloud &
  the 50 hardest HLT PF candidates within $\Delta R<0.5$ of the electron, plus up to 30 additional pixel tracks.
  13 features: $(p_T,\eta,\phi$, 5-way type one-hot, $d_z, d_{xy}, \chi^2/\mathrm{ndof})$,
  with $(\Delta\eta,\Delta\phi)$ relative to the electron \\
HLT input (regressed) &
  $E$, ECAL isolation, $\Delta\eta_{\rm in}$, $\Delta\phi_{\rm in}$,
  $\eta$, $\phi$ \\
HLT auxiliary &
  $p_T$, $\sigma_{i\eta i\eta}$, $H/E$, HCAL isolation,
  track isolation, normalized $\chi^2$, valid/lost/valid-pixel hit counts, quality bitmask, $d_0$,
  $d_z$, charge \\
Event auxiliary &
  event timestamp, $\rho$, nPV, nTracks \\
Offline target &
  $E$, ECAL isolation, $\Delta\eta_{\rm in}$, $\Delta\phi_{\rm in}$,
  $\eta$, $\phi$ \\
Additional offline information (not regressed) &
  $p_T = E/\cosh\eta$ \\
\bottomrule
\end{tabular}
\begin{tabular}{@{}p{3.4cm}p{9.8cm}@{}}
\toprule
Jets & Fields \\
\midrule
Particle cloud &
  the HLT jet's 50 hardest PF constituents, plus up to 30
  unpromoted pixel tracks, merged into one
  graph. 13 features: $(p_T,\eta,\phi$, 5-way type one-hot,
  $d_z, d_{xy}, \chi^2/\mathrm{ndof})$, with $(\Delta\eta,\Delta\phi)$
  relative to the jet axis \\
HLT input (regressed) &
  $p_T$, $\eta$, $\phi$, mass, charged EF, neutral hadron EF \\
Event auxiliary &
  event timestamp, $\rho$, nPV, nTracks \\
Offline target &
  $p_T$, $\eta$, $\phi$, mass, charged EF, neutral hadron EF \\
Additional offline information (not regressed) &
  photon EF $=1-{\rm charged}-{\rm neutral}$ \\
\bottomrule
\end{tabular}
\normalsize
\end{table}

\begin{figure}[p]
\centering
\includegraphics[width=0.98\textwidth]{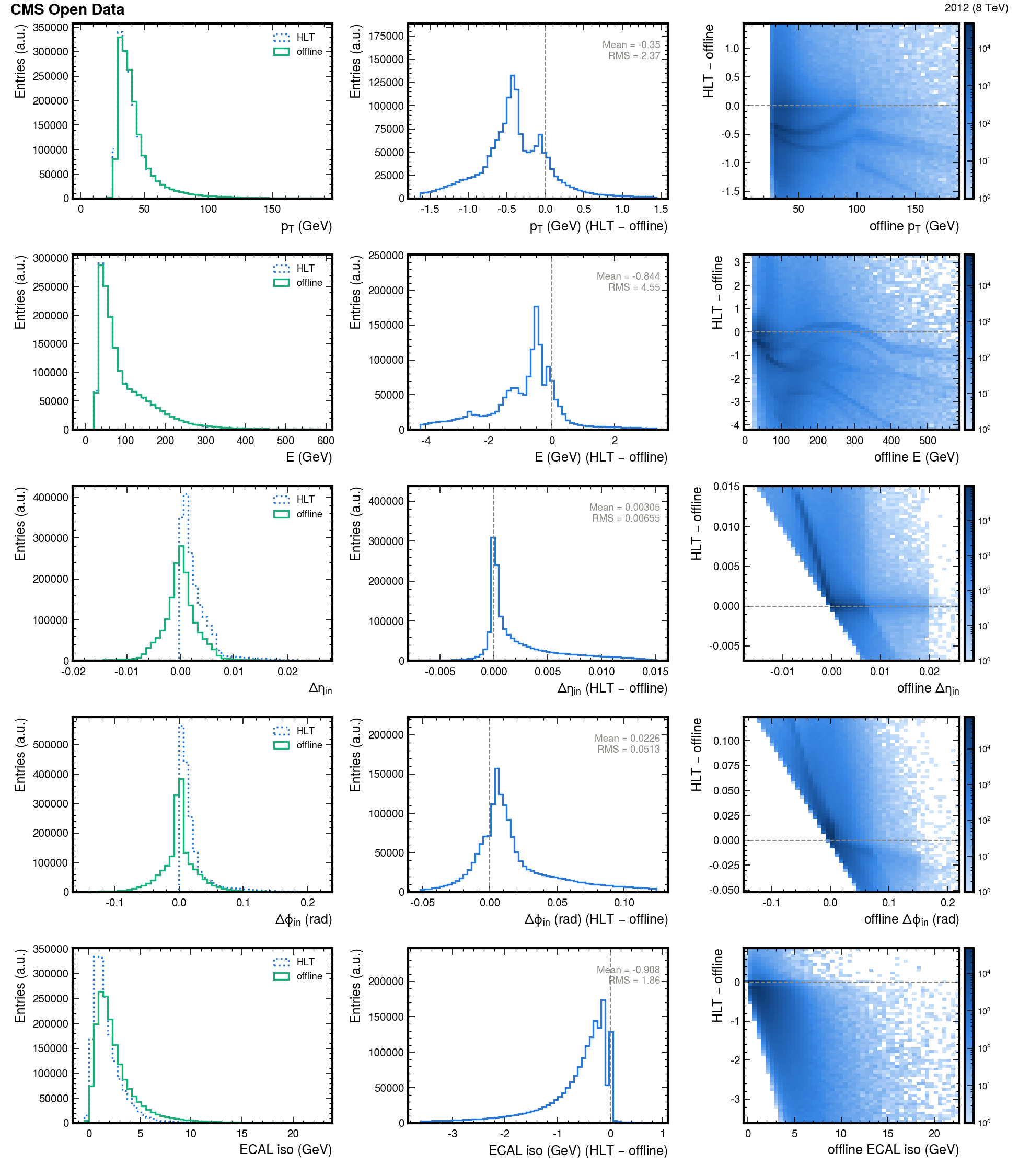}
\caption{Comparison between the HLT input and the offline target for
  electron-related quantities. Online and offline 1D distributions (left), raw
  HLT$-$offline residual (centre), and residual vs offline value (right) are
  shown for (from top to bottom):  $p_T$, $E$, $\delta\eta_{\rm in}$, $\delta\phi_{\rm in}$,
  and ECAL isolation.\label{fig:raw-res-ele}}
\end{figure}

\begin{figure}[p]
\centering
\includegraphics[width=0.98\textwidth]{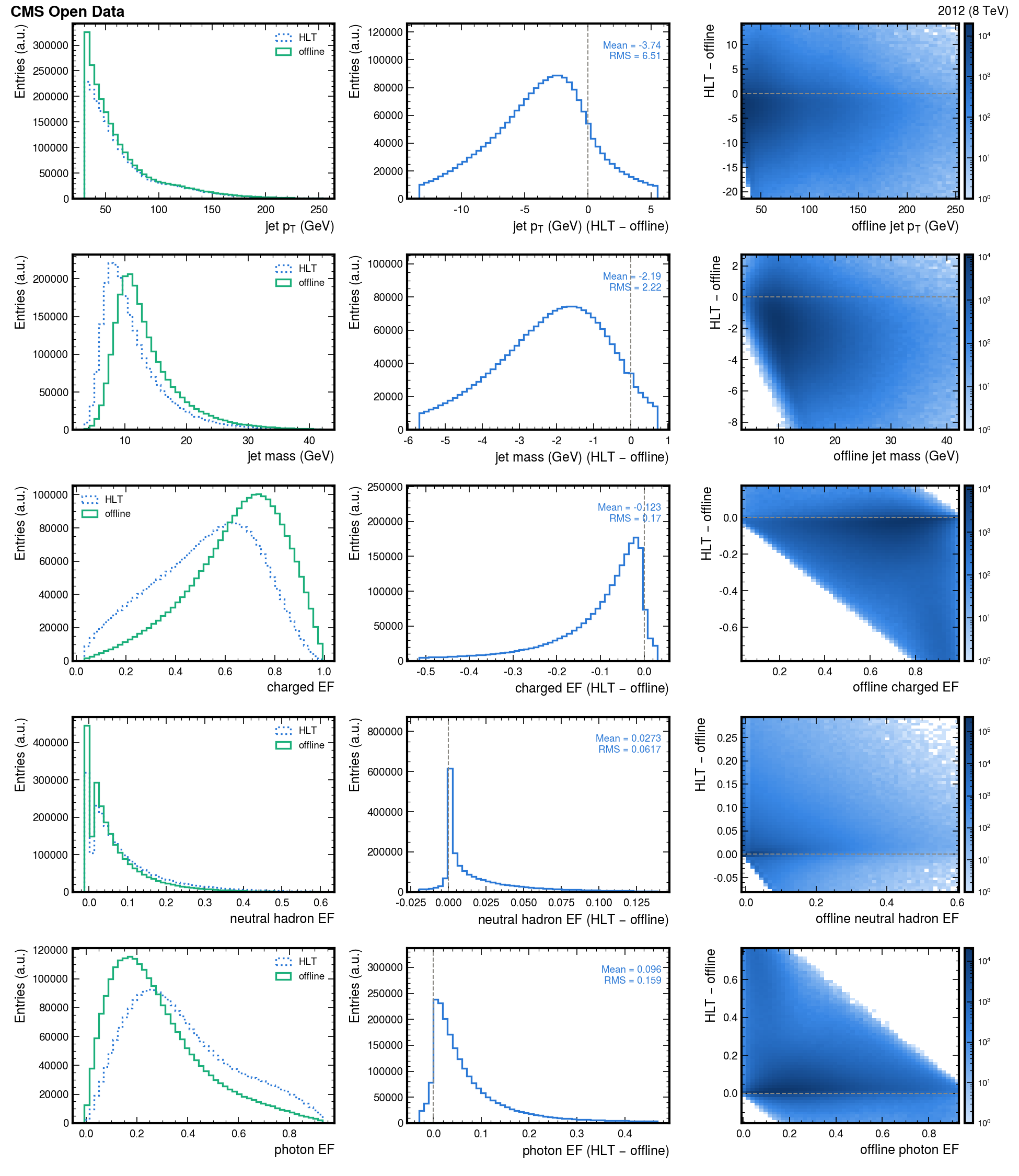}
\caption{Comparison between the HLT input and the offline target for
  jet-related quantities. Online and offline 1D distributions (left),
  raw HLT$-$offline residual (centre), and residual vs offline value
  (right) are shown for (from top to bottom): $p_T$, mass, charged,
  neutral, and photon energy fractions. \label{fig:raw-res-jet}}
\end{figure}

The main limitation when building the dataset is the use of the
offline global tag for the online conditions. Because of this, any
online electron quantity related to the calorimeter clustering is a
bit-exact match of the offline quantity, since the underlying
reconstruction algorithm is identical. This is why certain quantities
that one would normally regress (e.g., $\sigma_{i\eta i\eta}$ and
$H/E$) are excluded from the target (but still provided as inputs to
learn the conditional dependence of the targets on them, if any).
This unavoidable simplification does not diminish the relevance of our
results as proof of principle: a genuine online-to-offline difference
is introduced by the different tracking algorithm, with consequences
at various levels, as shown in
Figure~\ref{fig:raw-res-ele}. Nevertheless, one should keep in mind that
the actual problem one would face in the trigger would be more
complicated and it would require a higher-dimension regression. This
is less of a problem for the jet, where many online-to-offline
difference conspire to make the input and target quantities
substantially different, as shown in Figure~\ref{fig:raw-res-jet}.

\begin{table}[h]
\centering
\caption{Electron (top) and jet (bottom) yields and the
population surviving the event selection. \label{tab:yield}}
\begin{tabular}{@{}lr@{}}
\toprule
Stage & Count \\
\midrule
RAW events processed                                    & 1{,}868{,}142 \\
Events firing \texttt{HLT\_Ele27\_WP80\_v10} with $\geq1$ HLT electron & 1{,}352{,}356 (72\%) \\
HLT electron candidates                                 & 2{,}071{,}470 \\
Matched to an offline \texttt{gsfElectron}, $\Delta R<0.1$ & 1{,}832{,}757 (88\%) \\
After $p_T^{\rm HLT}>27$\,GeV and $p_T^{\rm off}\geq20$\,GeV & 1{,}748{,}331 \\
\bottomrule
\end{tabular}
\begin{tabular}{@{}lr@{}}
\toprule Stage & Count \\ \midrule RAW events processed &
1{,}868{,}142 \\ Events with $\geq1$ HLT jet passing the selection &
1{,}660{,}322 \\ HLT jet candidates & 4{,}250{,}092 \\ Matched to an
offline \texttt{ak5PFJet}, $\Delta R<0.4$ & 3{,}130{,}745 (74\%)
\\ After $p_T^{\rm off}\geq30$\,GeV and the energy-fraction closure
cut & 2{,}359{,}436 \\ \bottomrule
\end{tabular}
\end{table}

Table~\ref{tab:yield} give the resulting yields for the two object
types and the final dataset statistics. Each dataset is split object
by object, uniformly at random with a fixed seed, into training,
validation and test subsets in a 2:1:1 ratio, giving held-out test
samples of 532{,}509 jets and 396{,}159 electrons. The validation
split is used only for model selection, the retained checkpoint being
the one with the lowest validation loss over the course of training,
while the test split is never seen during training or model selection
and is used only for the results of
Section~\ref{sec:results}. Splitting per object rather than per event
is legitimate here because the detector response to two objects in the
same event is uncorrelated, which is the first-order approximation
every reconstruction algorithm rests on.

The high-level input fields are transformed before being presented to the
network, with the transform parameters determined on the training split
alone. The same transform is used for both objects:
\begin{itemize}
\item Feature values $x$ are replaced by $\log x$ for steeply falling,
  strictly positive quantities spanning orders of
magnitude ($p_T$, $E$, jet mass).
\item $\log(1+x)$  is used for isolation sums, positive defined but
  frequently exactly zero.
\item Azimuth is periodic and is replaced by its $(\sin\phi,\cos\phi)$
encoding, avoiding the spurious discontinuity a plain-angle input would
have near the $\pm\pi$ boundary.
\item The timestamp, the event-level pileup proxies and the HLT-only
track-quality variables are rescaled to the training split's $[0,1]$ range.
\end{itemize}
Every resulting column is then $z$-score standardized to zero mean and
unit variance. 

The regression \emph{targets} are not transformed at all: the
network's outputs, interpreted as online-to-offline residuals, are
summed to the corresponding HLT value directly in physical units and
the training loss is computed against the physical-unit offline truth.

\section{Regression Model and Loss Function}
\label{sec:model}\label{sec:kgc}\label{sec:loss-space}

The model is designed to consume an unordered, variable-length set of
particles alongside a fixed-length vector of high-level
quantities. Any point-cloud architecture satisfies this. We use a
ParticleNet-inspired graph regressor~\citep{qu2020particlenet}:
ParticleNet is well established for jet-constituent inputs in CMS,
compact enough to train quickly on samples of this size, and its
behaviour on particle clouds is well understood, which makes it a
reliable vehicle for measuring what the constituents contain.
Transformer-based set architectures~\citep{qu2022part} have since become
the stronger performer on comparable constituent-level tasks, and would be the
natural choice for a production deployment; we would expect them to
match, and probably improve on the results derived here. On the other
hand, the conclusions should not change significantly: what limits the
present results is in most cases the information content of the inputs
rather than the capacity of the model. 

\begin{figure}[t!]
\centering
\includegraphics[width=0.94\textwidth]{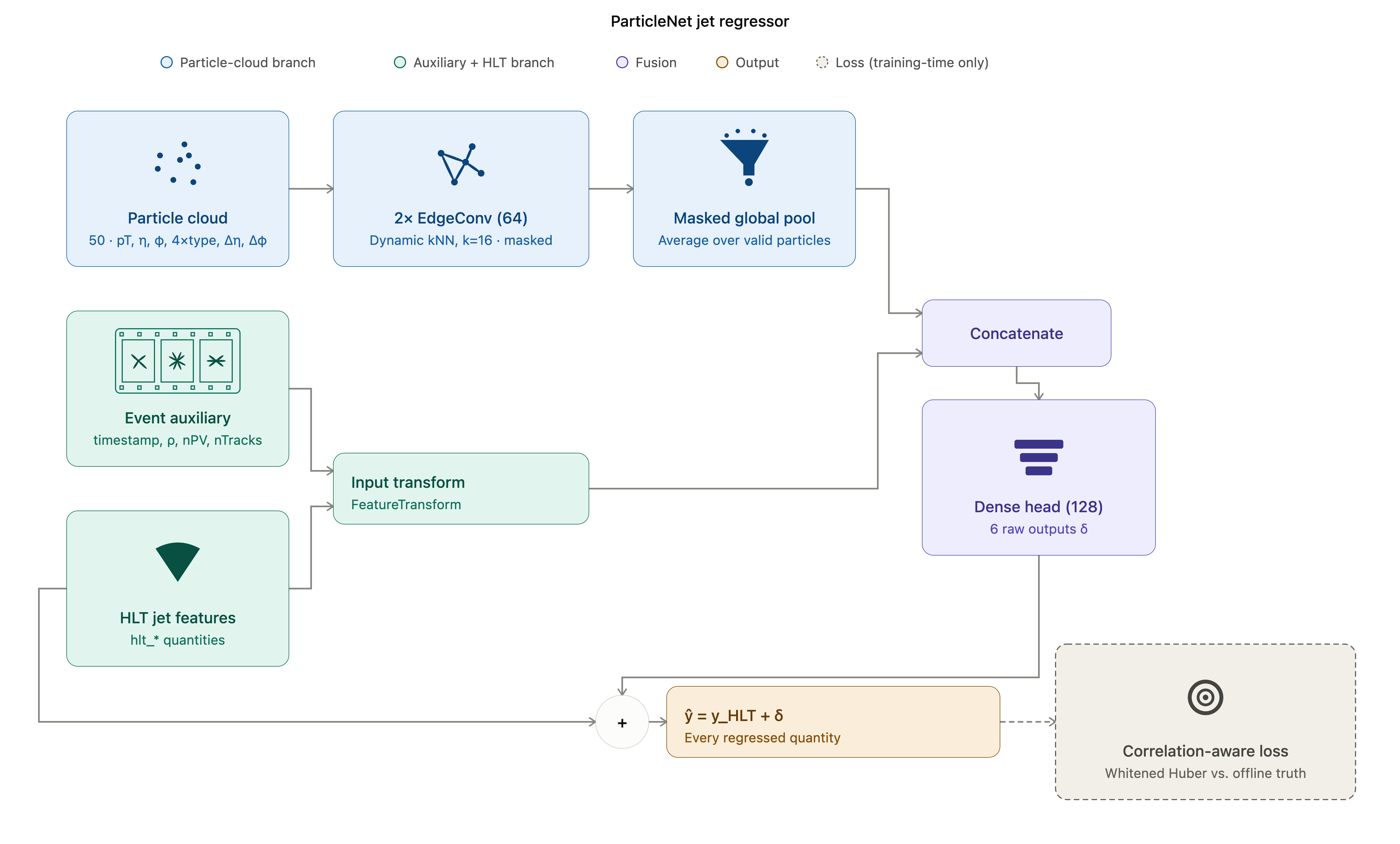}
\caption{Network architecture used for the jet and electron regression.
\label{fig:architecture}}
\end{figure}

The network architecture, the same for both objects, is shown in
Figure~\ref{fig:architecture}. It extends the particle cloud with the
high-level features as follows:
\begin{itemize}
\item The 80 candidate slots pass through a stack of two EdgeConv
  blocks \citep{wang2019dgcnn} of 64 channels each. Each block builds
  a $k$-nearest- neighbour graph with $k=16$ and, for every candidate
  $i$ and neighbour $j$, forms the edge feature $[x_i,\,x_j-x_i]$,
  passes it through a shared multilayer perceptron, and mean-aggregates over neighbours,
  with a linear shortcut. The graph is \emph{dynamic}: the first
  block's neighbours are found in physical $(\eta,\phi)$ space (with a
  proper $\phi$ wrap in the distance), while each later block rebuilds
  them in its own learned feature space, exactly as ParticleNet does.
\item Each candidate additionally carries its offset $(\Delta\eta,
  \Delta\phi)$ from the object it belongs to, with $\Delta\phi$
  wrapped to $[-\pi,\pi]$.
\item Padded slots are excluded everywhere through the validity mask:
  never chosen as neighbours, never contributing to an aggregation,
  zeroed as centres, and dropped from the pooling.
\item A masked global average pool gives one vector per object. The
  high-level features are \emph{concatenated} to that pooled vector
  and a 128-unit dense head with 10\% dropout produces the raw
  outputs, one per regressed quantity.
\end{itemize}

The raw network outputs are interpreted as residuals and summed to the
corresponding HLT values, in physical units: $\hat y_q = y_q^{\rm HLT}
+ \delta_q$.  The per-quantity residuals must then be combined into a
single loss, which is problematic whenever one term is bigger than the
others.  In early attempts with a simple mean, ECAL isolation
accounted for 72\% of the electron loss and the two energy fractions
83\% of the jet loss purely through scale. In such a configuration,
learning to regress the subdominant terms would be challenging.

To avoid this issue, we adopt a correlation-aware whitened Huber
loss, which removes this scale dependence and, in the same step,
accounts for correlations between residuals of different targets.

Each residual is first taken in the space that makes it a sensible error
measure. As a default, we use the plain residual except for
\begin{itemize}
\item a $\pi$-wrap for $\phi$, without which the few objects
  straddling $\pm\pi$ dominate the residual entirely;
\item for the strictly positive quantities of wide dynamic range ($E$
  and ECAL isolation) we use a $\log1p$ function
  $\mathrm{slog}(x)=\mathrm{sign}(x)\log(1+|x|)$, so that the residual
  behaves like a relative error while staying finite at zero and
  differentiable if the correction overshoots negative;
\end{itemize}

One can write the residual vector as:
\begin{equation}
r = \big(r_1,\ldots,r_Q\big)^{\mathsf T},
\qquad
r_q = f_q(\hat y_q) - f_q(y_q^{\rm off}),
\end{equation}
where $f_q$ is the identity or the signed $\log1p$ as above. For each term, we take as a
reference  its corresponding no-regression term ($\hat y \equiv y^{\rm HLT}$),
whose expectation value is denoted as $\langle\cdot\rangle_0$. We then divide
each term by the standard deviation of its no-regression term:
\begin{equation}
s_q = \sqrt{\big\langle (r_q - \langle r_q\rangle_0)^2\big\rangle_0},
\qquad
\tilde r_q = r_q / s_q ,
\qquad
{\rm Var}_0\big(\tilde r_q\big) = 1 \ \ \forall q ,
\label{eq:whiten-scale}
\end{equation}
so that every target has equal weight in the nothing-was-learned limit.
Because the $\tilde r_q$ then have unit variance, their covariance
\emph{is} their correlation matrix,
\begin{equation}
\Sigma_0 = {\rm Cov}_0\big(\tilde r\big),
\qquad
(\Sigma_0)_{qq'} = \rho\big(r_q, r_{q'}\big)\big|_0 .
\label{eq:whiten-cov}
\end{equation}
Table~\ref{tab:sigma0} gives the measured
$\Sigma_0$. They are not diagonal, and the off-diagonal entries are
driven by physical effects. For example, the correlation between the jet $p_T$
and mass residual is $+0.76$, since a jet whose charged constituents were not
tracked online loses momentum \emph{and} mass together. 

Factoring $\Sigma_0 = L_0 L_0^{\mathsf T}$ (Cholesky decomposition) and defining
\begin{equation}
y = L_0^{-1}\, \tilde r ,
\qquad\text{so that}\qquad
{\rm Cov}_0\big(y\big)
 = L_0^{-1}\,\Sigma_0\, L_0^{-\mathsf T} = \Id
\end{equation}
gives residuals that are uncorrelated and of unit variance at the
no-regression point. This is the right object to penalize because
\begin{equation}
\|y\|^2 \;=\; \tilde r^{\mathsf T}\,\Sigma_0^{-1}\,\tilde r ,
\label{eq:gls}
\end{equation}
i.e.\ squaring the whitened residual measures the original residual in the
$\Sigma_0^{-1}$ metric, which is generalized least squares: the network is
not penalized twice for one physical failure seen in two correlated
targets, while a prediction that breaks a correlation the data obey is
penalized more heavily.

The loss is finally
\begin{equation}
\mathcal{L} \;=\;
\frac{\big\langle H_\delta(y)\big\rangle}{\big\langle H_\delta(y)\big\rangle_0},
\qquad
H_\delta(y) = \frac{1}{Q}\sum_{q=1}^{Q}
\begin{cases}
\tfrac12 y_q^2 & |y_q|\le\delta ,\\[2pt]
\delta\big(|y_q|-\tfrac12\delta\big) & |y_q|>\delta ,
\end{cases}
\end{equation}
whose quadratic core uses the correlation structure of Eq.~\ref{eq:gls}
while the linear tail stops surviving outliers from taking over the
gradient; $\delta=1$ is the natural knee because $y$ is unit-variance at
the no-regression point by construction. Dividing by
$\langle H_\delta(y)\rangle_0$ fixes the interpretation of the loss:
$\mathcal{L} = 1$ for the no-regression configuration $\hat y = y^{\rm HLT}$
while  $\mathcal{L} < 1$ is the surviving fraction of the HLT residual.
To guarantee the $\mathcal{L} = 1$ no-regression condiction,
$\Sigma_0$ is pre-computed on the training dataset before starting the training.
One could renounce to having a reference scale for the training and instead
learn the $\Sigma$ matrix.

\begin{table}[h]
\centering
\caption{ The measured residual correlation matrix $\Sigma_0$ of
  Eq.~\ref{eq:whiten-cov} for electrons (top) and jets (bottom),
  evaluated on the training split at the no-regression point.  The
  matrix is symmetric by construction, so only the upper triangle is
  shown. The last row gives the per-quantity normalizations $s_q$ of
  Eq.~\ref{eq:whiten-scale}, in the units of the space each residual
  is measured in.
  (Section~\ref{sec:loss-space}).
  \label{tab:sigma0}}
\small
\begin{tabular}{@{}lrrrrrr@{}}
\toprule
Electron & $E$ & ECAL iso & $\Delta\eta_{\rm in}$ & $\Delta\phi_{\rm in}$ & $\eta$ & $\phi$ \\
\midrule
$E$                     & $1$ & $0.00$ & $0.01$ & $0.00$ & $0.00$ & $0.00$ \\
ECAL iso                &     & $1$ & $-0.15$ & $-0.10$ & $0.00$ & $0.00$ \\
$\Delta\eta_{\rm in}$   &     &     & $1$ & $0.15$ & $-0.23$ & $0.00$ \\
$\Delta\phi_{\rm in}$   &     &     &     & $1$ & $0.00$ & $0.11$ \\
$\eta$                  &     &     &     &     & $1$ & $0.01$ \\
$\phi$                  &     &     &     &     &     & $1$ \\
\midrule
$s_q$ & $0.0375$ & $0.2215$ & $0.00640$ & $0.0513$ & $0.00681$ & $0.0208$ \\
\bottomrule
\toprule
Jet  & $p_T$ & $\eta$ & $\phi$ & mass & charged EF & neutral EF \\
\midrule
$p_T$        & $1$ & $0.00$ & $0.00$ & $0.76$ & $0.16$ & $-0.13$ \\
$\eta$       &     & $1$ & $0.00$ & $0.00$ & $0.00$ & $0.01$ \\
$\phi$       &     &     & $1$ & $0.00$ & $0.00$ & $-0.01$ \\
mass         &     &     &     & $1$ & $0.10$ & $-0.03$ \\
charged EF   &     &     &     &     & $1$ & $-0.35$ \\
neutral EF   &     &     &     &     &     & $1$ \\
\midrule
$s_q$ & $6.46$\,GeV & $0.0358$ & $0.0391$ & $2.22$\,GeV & $0.1695$ & $0.0616$ \\
\bottomrule
\end{tabular}
\normalsize
\end{table}

\section{Results}
\label{sec:results}

Both networks are trained with the Adam optimizer~\citep{kingma2015adam}
for 100 epochs under a warmup-plus-cosine learning-rate schedule (a linear
ramp over the first 5 epochs, then a cosine decay to $\approx0$), with
gradient-norm clipping at 5. The checkpoint with the lowest validation
loss is retained. Training runs on a single GPU (NVIDIA V100),
using PyTorch~\citep{paszke2019pytorch}.

\begin{figure}[h]
\centering
\begin{minipage}{0.49\textwidth}
\centering
\includegraphics[width=\textwidth]{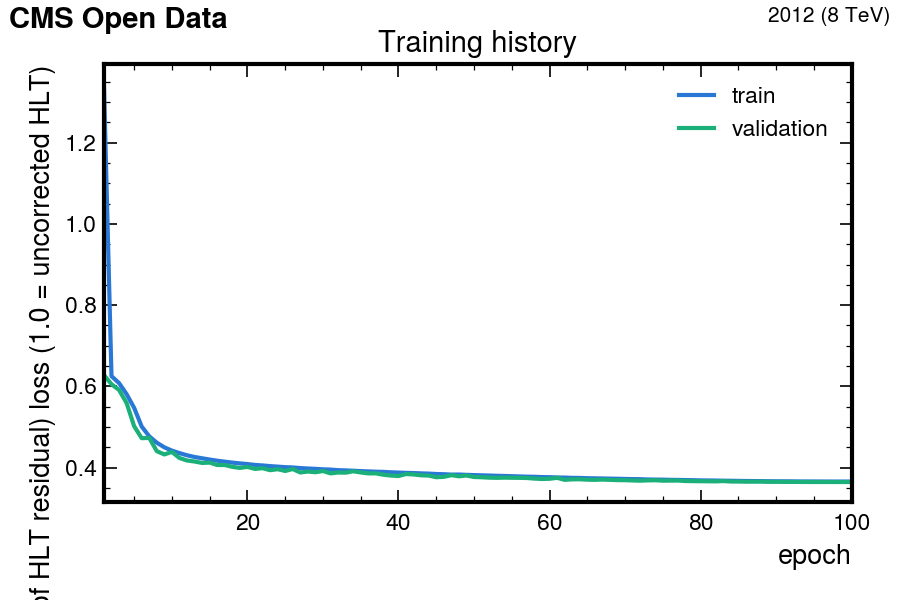}
\end{minipage}
\hfill
\begin{minipage}{0.49\textwidth}
\centering
\includegraphics[width=\textwidth]{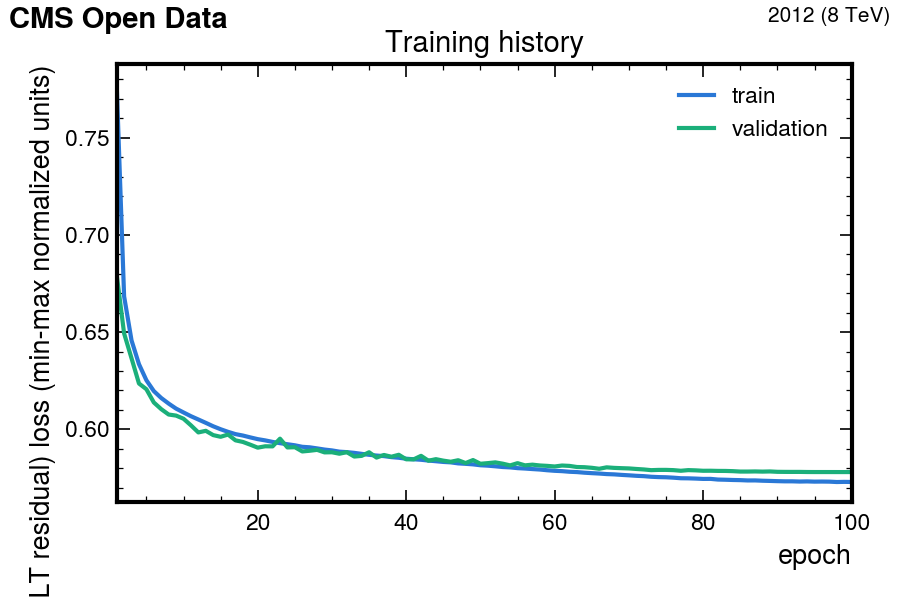}
\end{minipage}
\caption{Training and validation loss vs.\ epoch, for the electron 
(left) and the jet (right) models. \label{fig:history}}
\end{figure}

Figure~\ref{fig:history} shows the training and validation loss as a
function of epoch for both models. The validation curve is used to
select the best-epoch checkpoint. Neither model shows meaningful
overfitting. Most of the gain is made in the first 20--30 epochs.  The
electron validation loss falls to 0.365 and the jet one to 0.578: that
is, roughly 64\% of the electron and 42\% of the jet HLT residual is
removed by the correction, in the whitened metric averaged over each
object's six targets.

\subsection{Electron results}
\label{sec:res-electrons}

\begin{figure}[p]
\centering
\includegraphics[width=0.98\textwidth]{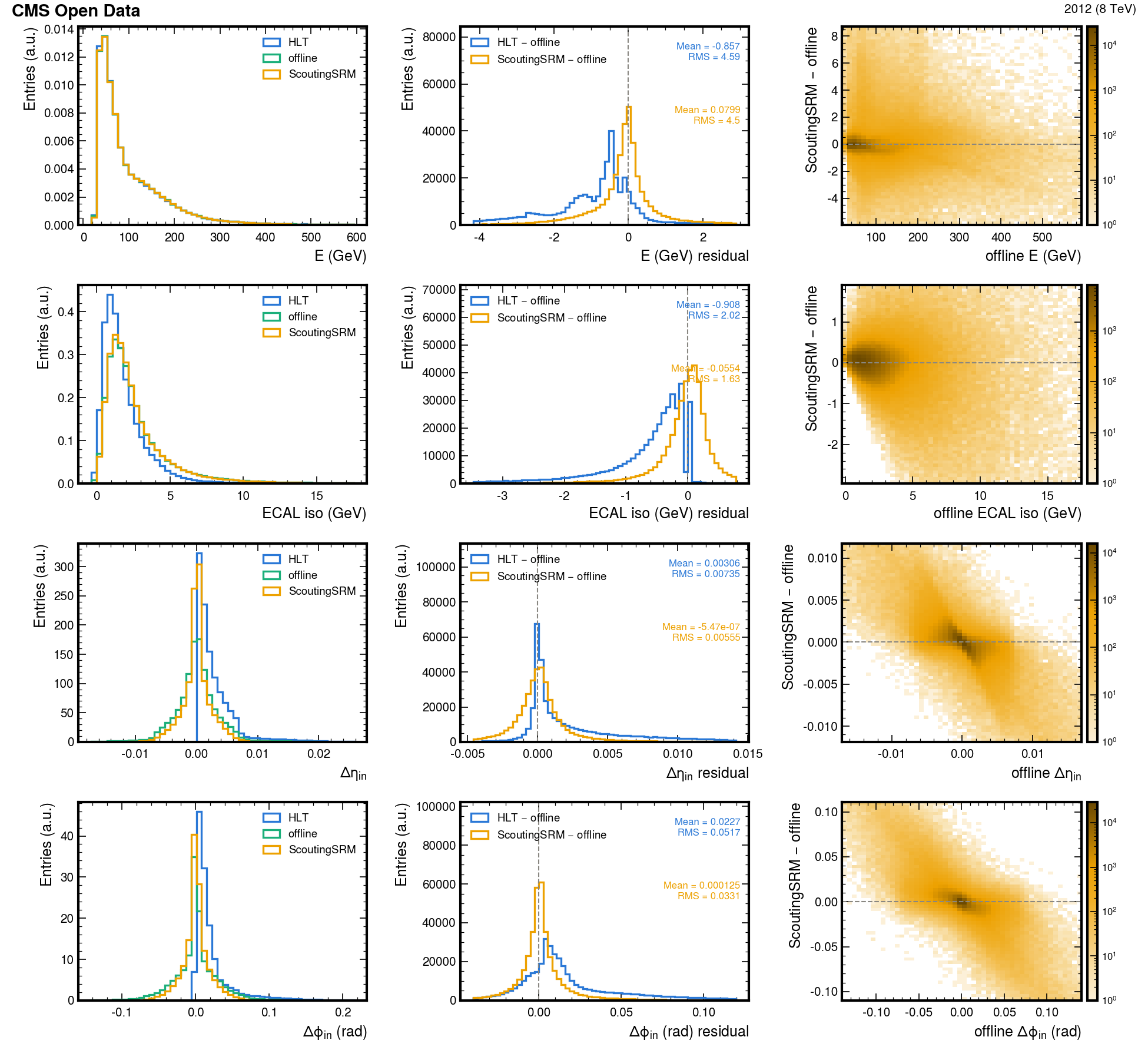}
\caption{\textbf{Electrons.} HLT, offline and ScoutingSRM-inferred
distributions (left), the residual before (HLT$-$offline) and after
(ScoutingSRM$-$offline) correction with mean and RMS annotated for both
(centre), and the corrected residual against the offline value (right).
The four quantities with a sizeable HLT/offline gap are shown; the
regressed $\eta$ and $\phi$ are omitted because the HLT already determines
them to $\approx1\%$ of the offline spread, so all three curves coincide
(their numbers are in Table~\ref{tab:res-ele}). Every panel is in physical
units; the $\phi$ residual is $\pi$-wrapped (Section~\ref{sec:loss-space}).
In the right-hand column a flat band on zero is the ideal: structure
there means the surviving error still depends on the true value.
\label{fig:res-ele}}
\end{figure}

Figure~\ref{fig:res-ele} shows the result of the electron regression,
comparing the network's inferred (``ScoutingSRM'') estimate to the
online input and the offline target. For every regressed quantity,
three plots are shown: the comparison of the three (HLT, offline, and
ScoutingSRM) distributions, the HLT-to-offline and
ScoutingSRM-to-offline residuals, and the two-dimensional distribution
of the ScoutingSRM-to-offline residual vs. the offline quantity.  The
figure only considers the four regressed electron quantities that
carry a sizeable HLT/offline difference. The other quantities are
already good to start with and the ScoutingSRM predictions confirm
that.  Forcing the network not to disrupt an already good HLT input
was the main motivation behind the choice to learn residuals and build
the prediction with a skip connection.

Table~\ref{tab:res-ele} quantifies the improvement provided by the
ScoutingSRM prediction, through the mean, the root mean square (RMS) and the mean absolute
value of the residual distribution.

The energy regression shows how these different ways to quantify the result
provide complementary information. Its bias is corrected from $-0.86$
to $+0.08$\,GeV and its mean absolute residual improves by 34\%, while
its RMS barely moves (2.0\%). Figure~\ref{fig:res-ele} shows why: the
corrected distribution is a sharp peak at zero replacing a broad,
negatively-offset one, but it retains a tail that the RMS, being a
mean-square statistic, weights far above the typical electron. In
fractional terms the resolution goes from 5.30\% to 5.27\%, while the
mean absolute fractional residual improves from 1.76\% to 1.19\%. The
correction fixes the energy \emph{scale} and the typical electron, but
it does not manage to narrow the tail. To do so, one would have to learn
$\Sigma_0$ instead of fixing it to its no-regression value, and turn the
output into a distributional regression, in which both the point estimate
and its uncertainty are learned as outputs. The price one would pay is giving up
the possibility of interpreting the loss value in absolute terms (a nice
but dispensable feature). The derived $p_T=E/\cosh\eta$ shows the same
behaviour, as it must.

The isolation and the track-matching variables improve on every statistic.
The systematic online underestimate of the ECAL isolation is removed to
the 6\% level of its own former size, and its mean absolute residual
improves more than that of any other quantity: this is what the particle
cloud was introduced for, since a scalar HLT isolation sum can only be
rescaled whereas the cone's candidates can be re-summed.
$\Delta\eta_{\rm in}$ and $\Delta\phi_{\rm in}$ improve by 24\% and 36\%
in RMS, with their biases dropping by more than two orders of magnitude.
These were the two quantities whose HLT value was worse than useless as an
estimator: the network is not rescaling them, it is replacing them.

\begin{table}[h]
\centering
\caption{\textbf{Electrons.} Residual before (HLT) vs.\ after (ScoutingSRM)
correction on the held-out test split of 396{,}159 electrons, in physical
units for every quantity. $\langle|r|\rangle$ is the mean absolute
residual, quoted alongside the RMS because the two disagree for the energy
for the reason discussed in the text. The derived $p_T=E/\cosh\eta$ is
listed for completeness.
\label{tab:res-ele}}
\small\setlength{\tabcolsep}{4pt}
\begin{tabular}{@{}lrrrrrrrr@{}}
\toprule
& \multicolumn{2}{c}{Mean} & \multicolumn{3}{c}{RMS} & \multicolumn{3}{c}{$\langle|r|\rangle$} \\
\cmidrule(lr){2-3}\cmidrule(lr){4-6}\cmidrule(lr){7-9}
Quantity & raw & corrected & raw & corrected & impr. & raw & corrected & impr. \\
\midrule
$E$ (GeV)                    & $-0.858$             & $0.080$              & 4.59     & 4.50     & 2.0\%  & 1.64     & 1.08     & 34\% \\
ECAL iso (GeV)               & $-0.909$             & $-0.055$             & 2.02     & 1.63     & 19\%   & 0.926    & 0.392    & 58\% \\
$\Delta\eta_{\rm in}$        & $3.06\times10^{-3}$  & $-5\times10^{-7}$    & 0.00735  & 0.00555  & 24\%   & 0.00368  & 0.00185  & 50\% \\
$\Delta\phi_{\rm in}$ (rad)  & $2.27\times10^{-2}$  & $1.3\times10^{-4}$   & 0.0517   & 0.0331   & 36\%   & 0.0322   & 0.0151   & 53\% \\
$\eta$                       & $1.3\times10^{-5}$   & $-1.1\times10^{-5}$  & 0.00702  & 0.00453  & 35\%   & 0.00212  & 0.00129  & 39\% \\
$\phi$ (rad)                 & $-5\times10^{-7}$    & $-3.9\times10^{-5}$  & 0.0211   & 0.0132   & 38\%   & 0.00780  & 0.00402  & 48\% \\
\midrule
$p_T$ (GeV, derived)         & $-0.357$             & $0.050$              & 2.37     & 2.36     & 0.7\%  & 0.749    & 0.528    & 30\% \\
\bottomrule
\end{tabular}
\normalsize
\end{table}

\subsection{Jet results}
\label{sec:res-jets}

\begin{figure}[p]
\centering
\includegraphics[width=0.90\textwidth]{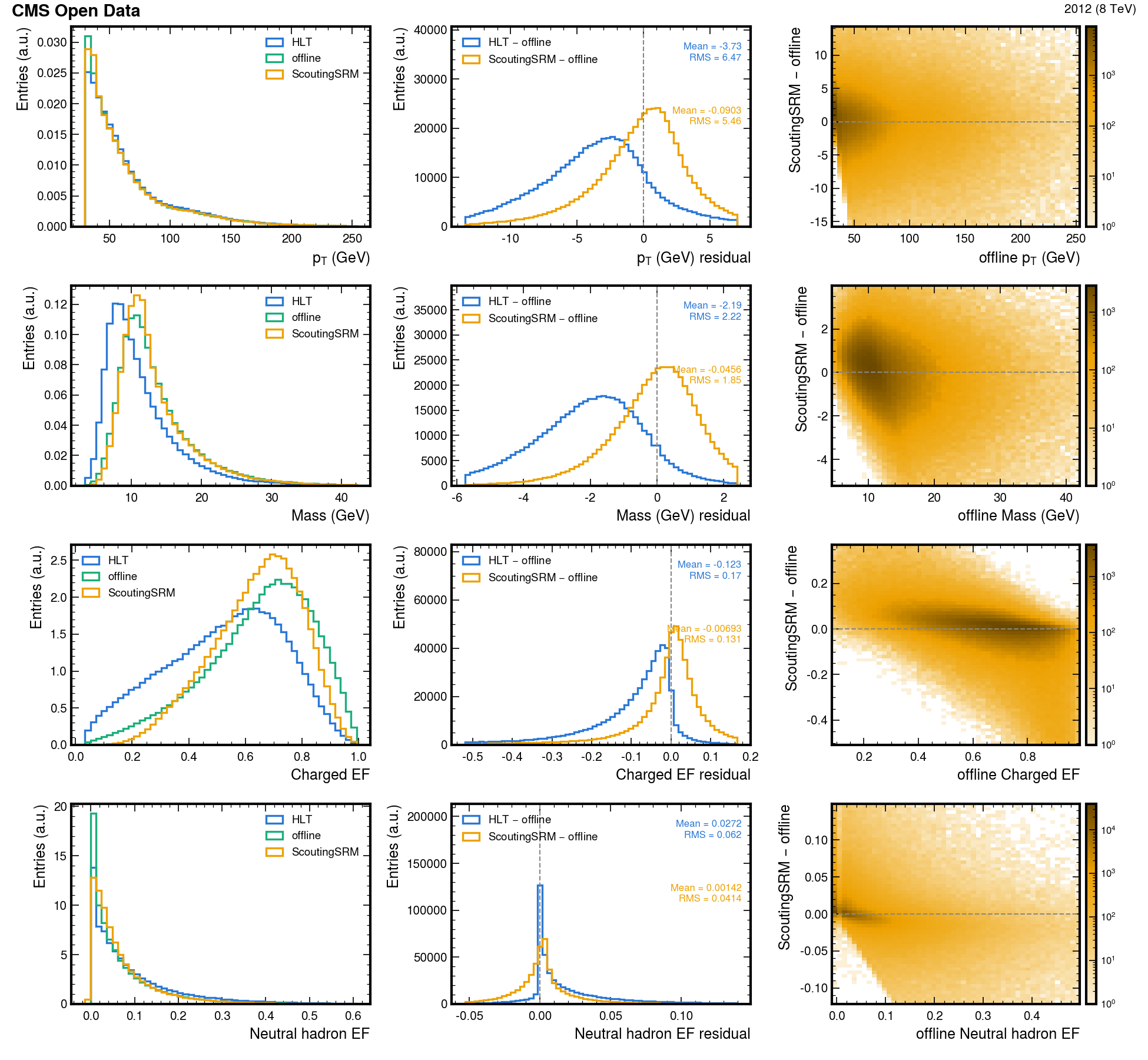}
\caption{\textbf{Jets.} The same as
Figure~\ref{fig:res-ele}. The $p_T$ axis starts at the 30\,GeV offline
selection. As for electrons, the regressed $\eta$ and $\phi$ are omitted: the HLT jet axis is already within 2--3\% of the offline spread, so the
corrected distribution simply reproduces the input one (Table
\ref{tab:res-jet}). In the composition and mass rows the right-hand panel
shows a sloped rather than flat band, regression towards the mean,
discussed in the text.
\label{fig:res-jet}}
\end{figure}

Figure~\ref{fig:res-jet} and Table~\ref{tab:res-jet} show the same
comparison for jets, in the same layout and the same physical units.

Every quantity improves in bias and in both spread statistics. The $p_T$
scale offset of $-3.73$\,GeV is removed almost completely, taking the
fractional resolution from 11.4\% to 8.5\%, and the mass offset collapses
from $-2.19$ to $-0.05$\,GeV. The composition improves throughout, with
the charged and neutral hadron fraction biases suppressed by roughly an
order of magnitude. The jet direction, expected to act as a null control
since the HLT already locates the axis to within a few per cent of the
offline spread, nonetheless improves by 9.2\% and 7.2\% in RMS.

The composition is the most instructive of these, because it is the
quantity for which the HLT value is least informative to begin with and
because the way it is recovered says what the particle cloud is actually
doing. All three online/offline differences of a jet share a single cause:
a charged particle whose track online tracking failed to reconstruct. The
particle does not disappear. Its energy is still deposited in the
calorimeters, but with no track to link to, particle flow reconstructs the
unlinked ECAL and HCAL clusters as a photon and a neutral hadron. The
energy is therefore relabelled, which is why the online jet is
$0.123$ lower in charged fraction and correspondingly higher in photon
($+0.096$) and neutral hadron ($+0.027$) fraction, the migration being
photon-dominated because a hadron deposits a substantial share of its
energy in ECAL. It is at the same time under-measured, because the
momentum estimate switches from the track, which the tracker determines to
the per-cent level, to a calorimetric measurement that under-responds in a
non-compensating calorimeter and carries the HCAL stochastic term. One
missing track thus produces the composition shift, the momentum and mass
deficit, and the degraded resolution together.

A correction acting on the HLT summary quantities alone cannot undo this,
and the particle cloud on its own cannot either. Every feature of the
particle-flow cloud is computed from the candidates the trigger already
built, so it can only restate what the trigger reported: measured on this
dataset, the cloud's own charged $p_T$ fraction correlates $0.998$ with the
HLT charged energy fraction and only $0.640$ with the offline one, which is
exactly the correlation the HLT value already has. What has to be
identified is which of the neutral deposits were charged particles to
begin with, and that information is absent from the candidates by
construction.

This is what the unpromoted pixel tracks supply. A pixel track with no
charged candidate attached to it marks the position of a charged particle
that the trigger saw at seed level and failed to promote. Its momentum is
of little use, since pixel-only tracking resolves $p_T$ poorly, but its
direction is well determined, and direction is what is needed: merged into
the same graph as the particle-flow candidates and built in
$(\eta,\phi)$, it places the evidence of a lost track in the same
neighbourhood as the neutral deposits that lost track produced. The
network can then learn that a neutral sitting next to an unpromoted pixel
track was a charged particle to begin with, and move both the energy and
the label back where they belong. What it performs is a statistical
re-attribution conditioned on the local topology rather than a
per-particle repair, which is also why the merge into a single graph
matters: two separate branches would discard precisely the adjacency the
inference rests on.

The residual-versus-truth panels of Figure~\ref{fig:res-jet} add a
diagnosis that the residual distribution alone cannot give. For the
composition and the mass the corrected residual is not a flat band on zero
but slopes downward, positive at low values of the offline quantity and
negative at high ones. One might then think that the improvement quoted
above is an artefact of the regression pushing the residuals to zero,
shrinking every prediction towards the population mean rather than
learning anything about the individual object. This is not the case, as
one can see once the tradeoff between bias and variance is made explicit (see
Table~\ref{tab:response}). To do so, one can write the reconstructed value as
$\hat y = a + b\,t + \varepsilon$ for a true value $t$: the slope $b$ is
the \emph{response} and $\sigma(\varepsilon)$ the scatter about it. The
residual RMS mixes the two,
\begin{equation}
\mathrm{Var}(\hat y - t) = (b-1)^2\,\mathrm{Var}(t) + \sigma^2 ,
\label{eq:response-decomp}
\end{equation}
so a model with a deliberately flat response is rewarded with a smaller
RMS without being a better estimator. The scatter $\sigma$, on the other
hand, is measured about the estimator's own best linear relation to the
truth, and therefore cannot be reduced by attenuating the response. In
Table~\ref{tab:response}, $\sigma$ falls for every jet quantity, by 16\% to
37\%: the corrected estimate genuinely carries more information than the
HLT one, and the improvement is not shrinkage.

The attenuation is nonetheless real and worth stating. The HLT is already
far from unit response on the charged energy fraction ($b=0.70$), and the
correction lowers it further to $0.60$ while reducing the scatter from
0.160 to 0.109. This is expected rather than a defect: the loss is
minimised by the conditional mean $E[t\,|\,x]$, which is attenuated
whenever the inputs underdetermine the target, so minimum variance and
unit response cannot be had simultaneously. Which of the two is wanted
depends on the use. For a per-object estimate the current behaviour is
optimal; for an analysis binning in the corrected quantity the attenuation
is a systematic, removable by a one-parameter rescaling exactly as jet
energy corrections are derived by measuring a response and inverting it.
The mass is the one quantity whose response the correction moves
\emph{towards} unity while also sharpening it, and the charged fraction
the opposite extreme, where the constituents leave the most ambiguity and
the estimator hedges hardest.

\begin{table}[h]
\centering
\caption{\textbf{Jets.} Residual before (HLT) vs.\ after (ScoutingSRM)
correction on the held-out test split of 532{,}509 jets, in the same
layout and the same physical-unit convention as Table~\ref{tab:res-ele}.
The photon energy fraction is omitted throughout: it is exactly
$1-{\rm charged}-{\rm neutral}$, so its residual is minus the sum of the
two composition rows. $\eta$ and $\phi$ are regressed but not plotted
(Figure~\ref{fig:res-jet}); their rows are given here for completeness.
\label{tab:res-jet}}
\small\setlength{\tabcolsep}{4pt}
\begin{tabular}{@{}lrrrrrrrr@{}}
\toprule
& \multicolumn{2}{c}{Mean} & \multicolumn{3}{c}{RMS} & \multicolumn{3}{c}{$\langle|r|\rangle$} \\
\cmidrule(lr){2-3}\cmidrule(lr){4-6}\cmidrule(lr){7-9}
Quantity & raw & corrected & raw & corrected & impr. & raw & corrected & impr. \\
\midrule
$p_T$ (GeV)          & $-3.73$              & $-0.090$             & 6.47   & 5.46   & 16\%     & 5.35   & 3.34   & 38\% \\
mass (GeV)           & $-2.19$              & $-0.046$             & 2.22   & 1.85   & 17\%     & 2.41   & 1.18   & 51\% \\
charged EF           & $-0.123$             & $-6.9\times10^{-3}$  & 0.170  & 0.131  & 23\%     & 0.134  & 0.0779 & 42\% \\
neutral hadron EF    & $2.73\times10^{-2}$  & $1.4\times10^{-3}$   & 0.0620 & 0.0415 & 33\%     & 0.0335 & 0.0215 & 36\% \\
\midrule
$\eta$               & $5.6\times10^{-5}$   & $1.5\times10^{-5}$   & 0.0359 & 0.0326 & 9.2\%    & 0.0190 & 0.0171 & 10\% \\
$\phi$ (rad)         & $-2.9\times10^{-4}$  & $4.8\times10^{-5}$   & 0.0392 & 0.0364 & 7.2\%    & 0.0216 & 0.0201 & 7.0\% \\
\bottomrule
\end{tabular}
\normalsize
\end{table}

\begin{table}[h]
\centering
\caption{\textbf{Jets.} Response $b$ and scatter $\sigma$ of
Eq.~\ref{eq:response-decomp}, for the uncorrected HLT value and the
correction, on the held-out test split. $\sigma$ falls for every quantity.
The composition improves in scatter while giving up response, which is the
attenuation discussed in the text; the mass is the one quantity whose response the
correction moves \emph{towards} unity while also sharpening it.
\label{tab:response}}
\small\setlength{\tabcolsep}{6pt}
\begin{tabular}{@{}lrrrrr@{}}
\toprule
& \multicolumn{2}{c}{HLT} & \multicolumn{2}{c}{ScoutingSRM} & \\
\cmidrule(lr){2-3}\cmidrule(lr){4-5}
Quantity & $b$ & $\sigma$ & $b$ & $\sigma$ & $\sigma$ gain \\
\midrule
$p_T$ (GeV)       & 0.990 & 6.45   & 0.983 & 5.42   & 16\% \\
mass (GeV)        & 0.881 & 2.10   & 0.914 & 1.77   & 16\% \\
charged EF        & 0.700 & 0.160  & 0.606 & 0.107  & 33\% \\
neutral hadron EF & 1.066 & 0.0617 & 0.831 & 0.0386 & 37\% \\
\bottomrule
\end{tabular}
\normalsize
\end{table}

\subsection{Response and resolution versus $p_T$}
\label{sec:res-vs-pt}

\begin{figure}[p]
\centering
\includegraphics[width=\textwidth]{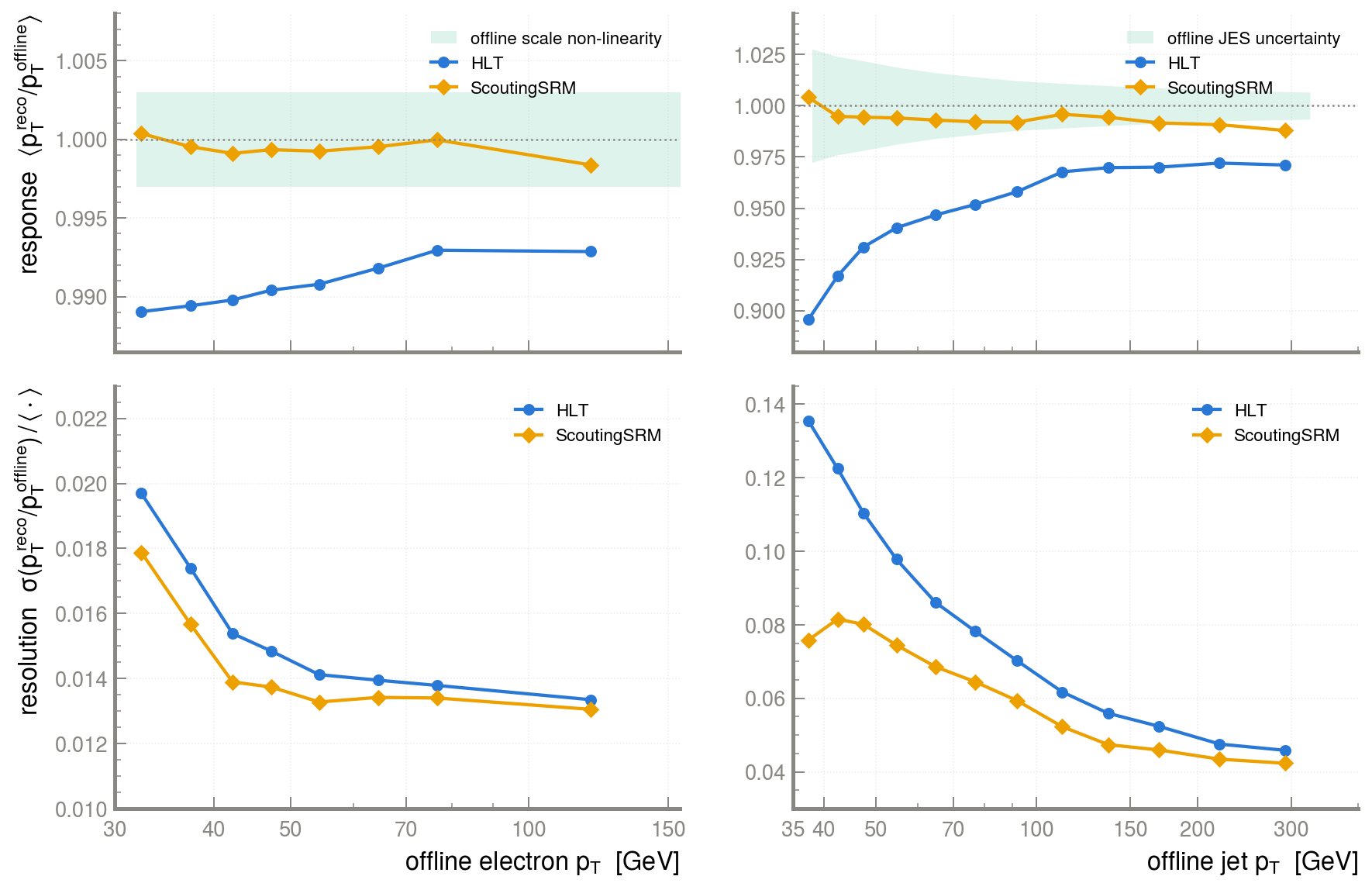}
\caption{Response (top) and resolution (bottom) against the offline $p_T$,
for electrons (left) and jets (right), comparing the uncorrected HLT object
with the ScoutingSRM one. The resolution is the width of
$p_T^{\rm reco}/p_T^{\rm offline}$ divided by its mean, so the two rows are
independent, with the outer 0.5\% trimmed from each side before the width is
taken. The green band on the jet response is the total Summer13\_V4 AK5PF
jet energy scale uncertainty~\citep{cms:jes8tev} from this dataset's own
conditions, averaged in
each bin over the $(p_T,\eta)$ distribution of the jets it contains. The
band on the electron response is the $0.3$\% electron energy-scale
calibration uncertainty CMS quotes for this dataset~\citep{cms:electron8tev}. 
Note the very different vertical ranges of the two columns: the HLT
electron energy starts an order of magnitude closer to offline than the HLT
jet $p_T$ does, and the contrast is the point of the side-by-side. Electrons
start at 30\,GeV and jets at 35\,GeV, one bin above the respective sample
edges: the lowest jet bin abuts the $p_T^{\rm offline}>30$\,GeV training
selection, which the network has learnt and which biases its response up and
narrows its width artificially, and the lowest electron bin is the trigger
turn-on, where only up-fluctuated objects are matched. Neither says anything
about the reconstruction.
\label{fig:resolution}}
\end{figure}

The residual statistics above are inclusive, and none of them shows how the
correction behaves as a function of the kinematics, which is where a
trigger-level method has to be judged: a scouting analysis lives near its
thresholds, and a correction that only worked on the objects already
measured well would not be worth deploying. Figure~\ref{fig:resolution}
therefore shows both objects the way reconstruction performance is
conventionally presented, in bins of the offline $p_T$: the mean and the
width of $p_T^{\rm reco}/p_T^{\rm offline}$, binned in the truth so that
migration stays out of the comparison, with the width quoted after dividing
out the response so that neither curve is rewarded for being attenuated.

For jets the correction acts as a $p_T$-dependent energy scale and a
resolution improvement at once. The HLT response is the familiar rising
curve of an uncalibrated jet, $10$\% low at 40\,GeV and recovering to
$2.9$\% low above 200\,GeV; the correction flattens it to within $1$\% of
unity, so what is removed is not an offset but its $p_T$ dependence. The
resolution improves most where the HLT is worst, from $13.6$--$9.8$\% to
$7.6$--$7.5$\% over 35--55\,GeV, roughly a factor $1.8$, and the two curves
converge by 200\,GeV: the method is worth the most exactly where a scouting
analysis is threshold-limited. For electrons the starting point is already
close to offline and the gain is correspondingly smaller, the correction
moving the response from $0.989$ to $1.000$ at 30\,GeV and holding it within
$0.2$\% of unity throughout.

The bands are the point of the figure. Deriving dedicated energy
corrections for trigger-level objects has been one of the practical
obstacles to using scouting data for measurements, and what the comparison
tests is not whether the correction is perfect but whether the corrected
object sits close enough to the offline one that the standard, centrally
provided calibration can simply be applied to it. It does: over most of the
spectrum the corrected jet lies inside the jet energy scale
uncertainty~\citep{cms:jes8tev} and
the corrected electron inside the $0.3$\% electron energy-scale
calibration~\citep{cms:electron8tev}, while the uncorrected HLT objects lie
outside both.

The band is the uncertainty on the correction from the reconstructed
to the true jet, that is, on the offline response itself. The
corrected object supplies a second estimate of that same response, so
wherever the two agree to better than the band they are equally good
estimates of the truth, and their difference is not an additional
systematic: the ignorance it would represent is already contained in
the band. This holds over most of the spectrum, and fails only in the
high-$p_T$ tail.\footnote{The deterioration at large $p_T$ is not
observed when training on the first half of the dataset, characterized
by a lower average pileup. This indicates a residual pileup
dependence, not captured by the auxiliary pileup proxies given as
input. The problem would be mitigated on more recent data, where
pileup-per-particle identification (PUPPI)~\citep{PUPPI} subtraction is
applied to both the offline and the scouting data, as part of the
post-processing steps preceding the regression inference.}

Two caveats belong with that statement. Both the online and
the offline jets here are \emph{uncorrected} PF jets, so the comparison is a
reconstruction closure and the jet energy corrections would be applied
identically on top of either. And the bands are scale uncertainties: they
say nothing about resolution, and the corrected jet remains appreciably
wider than an offline jet.

That residual width is worth weighing against the resolution an analysis
carries in any case. Because the part of the measurement common to the two
reconstructions cancels in the ratio, the width in
Figure~\ref{fig:resolution} is the \emph{additional} smearing that using a
scouting jet introduces on top of an offline one, and the two add in
quadrature. The offline jet energy resolution is itself 15--20\% at
30\,GeV and about 10\% at 100\,GeV~\citep{cms:jes8tev}; taking the lower
end, 15\%, as the conservative choice near threshold, an analysis at
35\,GeV would work with an effective resolution of
$\simeq20$\% using the uncorrected HLT jet and
$\simeq17$\% using the corrected one, against 15\%
offline. The correction thus reduces the resolution penalty for working
with scouting jets rather than offline ones from about a third to about a
tenth of the offline value.

The same weighing carries less force for electrons, whose offline energy
resolution is already at the per-cent level: there the online limitation
was never the width but the response, and it is the response that the
correction removes.

\section{Envisioned deployment}
\label{sec:deployment}

The regression demonstrated here is deliberately framed so that it
could be deployed without redesigning the online scouting
processing. The HLT continues to reconstruct and persist exactly the
same compact, high-level scouting objects it does today, at the same
rate and the same per-event bandwidth cost: nothing about the online
system changes, and no additional latency or online compute is
introduced. The regression step would instead sit entirely downstream,
as an offline post-processing stage applied after the scouting stream
has already been written: for each persisted object, a trained
regressor maps its HLT-level features and its persisted PF candidates
onto an inferred offline-quality estimate, before (or as part of)
producing the final analysis-ready dataset. Because this step consumes
only what the scouting stream already contains, it requires no access
to raw detector data.

The training of the regression models rely only on data and can be
performed periodically as detector conditions and calibrations evolve
over a run, independently of the online trigger menu.

The training sample is provided by the scouting monitoring stream, a
randomly sampled subset of scouting events additionally reconstructed
offline. On these data, every object carries both the
HLT and the offline view of its features, under the real detector conditions
and the real calibration lag the correction would be applied to.

This approach could be framed as a re-processing of the already
collected scouting data, but also as a real-time offline processing of
the scouting data to be collected during the High-Luminosity LHC (HL-LHC) phase.

\section{Conclusion}
\label{sec:conclusion}

This paper provides a demonstrator for offline-quality
super-resolution of scouting-level objects on CMS 2012 Open Data.  For
both electrons and jets, one common architecture taking as input the HLT
view of the object, the particle cloud around it and a few auxiliary
quantities is capable to regress the corresponding offline
view. Residual distributions show much smaller biases and a reduced
variance. In many cases, strong asymmetries and long tails are largely
suppressed, providing a much better behaving response. Notably, the
residual momentum non-closure is smaller than the uncertainty on the
offline momentum scale itself, making the online and offline quantities
much more similar for practical purposes. Also, the poorly reconstructed online jet mass is
improved to match very closely the offline one.

What the network exploits is seen most clearly in the jet composition. A
charged particle whose track the online reconstruction fails to build does
not disappear: its energy still reaches the calorimeters, but with no track
to link to, particle flow labels it a photon or a neutral hadron and
measures it calorimetrically rather than from the track. A single missing
track therefore produces the composition shift, the momentum and mass
deficit, and the resolution loss at once. The particle-flow candidates on
their own cannot undo this, since they can only restate what the trigger
reconstructed; the unpromoted pixel tracks can, because they mark where the
lost charged particles went, and placing them in the same graph as the
candidates lets the network associate a neutral deposit with the track that
should have produced it. In this sense the method does not invent
resolution: it recovers information that the online reconstruction did
produce and then discarded.

The gain is smallest on the quantities the HLT already determines
well: the jet axis, which the two views agree on to within a few per
cent of the offline spread, improves by only 7--9\% in RMS, against
16--33\% for the kinematics and composition. The electron energy
improves in bias and in typical case (34\% in mean absolute
residual). Its RMS barely moves, because the surviving error is
concentrated in a tail that a mean-square statistic weights far above
the typical electron.

One limitation bounds how far these numbers should be extrapolated. The
online conditions of the 2012 HLT menu are not published with the Open
Data, so the HLT replay must be run with the offline global tag. This
removes by construction the calibration-driven part of the online/offline
difference, and with it the very effect that motivates the electron case:
the drift of the ECAL response between the constants available at trigger
time and those derived afterwards. What is demonstrated here is the
recovery of a structural, algorithmic difference. There is reason to expect
the calibration term to be no harder: it is a smooth function of time and
detector position, and nothing is destroyed by a stale calibration
constant, the energy being measured but misscaled, whereas the tracking
difference corrected here requires inferring objects the online
reconstruction never built. The event timestamp is carried among the
network inputs for exactly this purpose, though in the present sample it is
necessarily inert, the shared global tag leaving no time dependence to
learn. Establishing this, however, requires paired
data taken under real online conditions, which is what the scouting
monitoring stream provides and this demonstrator cannot.

What this demonstrator establishes is that the ingredients for a deployment
are already in place. The scouting monitoring stream supplies, for a
randomly sampled subset of scouting events, both the HLT and the offline
view of every object under real detector conditions and real calibration
lag, so training demands no change to the trigger, to the recorded event
content, or to the online latency budget; and because the correction
consumes only what the scouting stream already contains, it can be applied
to the data already on tape as much as to data still to be collected.

Turning this into a usable dataset is, however, a substantial
undertaking rather than an extension of the present work. The two
object types treated here are a proof of concept: the same exercise
has to be carried out for photons, muons and taus, and possibly at the
level of the tracks and neutral hadrons from which composite objects
are built, each with its own feature schema, its own targets and its
own validation, and each under real online conditions, where the
calibration term absent from this demonstrator is present. On top of
that sits a second stage, for the event-level quantities that no
per-object correction can reach: the missing transverse momentum, for
instance, would be regressed from a point cloud of corrected objects
together with the global observables of the event, conditional on the
object-level corrections already applied. Building and validating such
a chain is the work of a dedicated team, and it is what would turn the
result reported here into a scouting dataset suitable to carry out a
broad-scope program, including final states not yet considered 
(e.g., with electrons and photons).

\section*{Acknowledgements}
Claude (Anthropic) was used as an AI coding, reasoning and writing
assistant throughout this work. Prompt with instructions on what to
do, it developed the reprocessing and matching pipeline, carried out
the regression training, and co-wrote the final manuscript. 

\bibliographystyle{unsrtnat}
\bibliography{refs}

\end{document}